\documentclass{JHEP3} % 10pt is ignored!

\usepackage{epsfig,multicol,bbm,amsmath}
\usepackage{graphicx}
\usepackage{amssymb}
\usepackage{mathrsfs}

\newcommand\beq{\begin{equation}}
\newcommand\eeq{\end{equation}}
\newcommand{\bea}{\begin{eqnarray}}
\newcommand{\eea}{\end{eqnarray}\noindent}

\graphicspath{{fig/}}

\title{Chiral Modulations in Curved Space II: Conifold Geometries}

\author{Antonino Flachi \\
        Multidisciplinary Center for Astrophysics, Instituto Superior T\'ecnico, Lisbon, Portugal\\
        E-mail: \email{antonino.flachi@ist.utl.pt}}

\abstract{In this paper, we extend our previous analysis concerning the formation of inhomogeneous condensates in strongly-coupled fermion effective field theories on curved spaces and include the case of conifold geometries that represent the simplest tractable case of manifolds with curvature singularities. In the set-up considered here, by keeping the genuine thermodynamical temperature constant, we may single out the role that curvature effects play on the breaking/restoration of chiral symmetry and on the appearance of inhomogeneous phases. The first goal of this paper is to construct a general expression of the finite temperature effective action for inhomogeneous condensates in the case of four-fermion effective field theories on conifold geometries with generic Riemannian smooth base (ge\-neralised cones). 
The other goal is to implement numerically the above formal results and construct self-consistent solutions for the condensate. We explicitly show that the condensate assumes a kink-like profile, vanishing at the singularity that is surrounded by a bubble of restored chiral symmetry phase.}


\begin{document}

\section{Introduction}

Current understanding of the phase structure of strongly interacting field theories is still far from being complete, despite the fact that this problem has been at the center of active interest for many years. 
The case of QCD is of particular physical relevance and many ongoing efforts are directed at mapping the geography of the phase diagram of the theory at finite temperature and density, understand the nature of the transitions between the different phases, locate the critical points, describe the morphology of the ground state. 
Impediments to arrive at the desired complete characterization of the phase structure of QCD are mainly due to the notorious difficulties in performing {\it ab initio} lattice computations %at finite density or 
under the influence of generic external conditions and, amongst the various approaches, the use of effective models is one of the most common playgrounds where the phase structure 
%chiral and confinement/deconfinement phase transitions 
can be discussed (see Refs.~\cite{review1,review2,review3,review4,review5,review6} for review). Within these effective models, the Nambu-Jona Lasinio (with its variants) is, possibly, the field theoretical set-up that received greatest attention. 

One particular aspect that has been at the center of recent discussions is the identification of inhomogeneous phases. In fact, the appearance of modulations in strongly-coupled fermionic systems has led to several new insights concerning inhomogeneous phases in QCD. More precisely, several works indicated that, in the chiral limit, inhomogeneous phases are energetically preferred for relatively large values of the chemical potential, and the preferred inhomogeneous ground state in the vicinity of the chiral critical point assumes a crystalline structure similarly to superconductors. The problem has been analyzed in different models and seems to be quite generic (see, for example, \cite{nakano,nickel1}). %In the Nambu-Jona Lasinio and in the Quark Meson models, the problem has been also addressed quantitatively in Ref.~\cite{nickel}. 
The inclusion of gauge degrees of freedom has also been discussed within models of the Nambu-Jona Lasinio class with the aid of vector interactions and Polyakov loops in Ref.~\cite{buballa2}.

The occurrence of inhomogeneous phases has also been studied in a number of different contexts including superconductivity (see for example Refs.~\cite{casalbuoni,alford,bowers,mannarelli,rajagopal,buballa,Casalbuoni:2005zp}), lower dimensional field theories and, in particular, the Gross-Neveu model with a large number of fermions (see \cite{basar1,basar2,basar3,schnetz,urlichs,boehmer,dgkl}) that allows for exact integrability. Earlier studies regarding inhomogeneous phases were also carried out in the context of the Skyrme model \cite{skyrme1,skyrme2,skyrme3,skyrme4,skyrme5} and in four-fermion models \cite{4fermi}. 

Strongly coupled theories are physically relevant also in applications of astrophysical and cosmological nature (neutron stars, black holes, cosmological phase transitions). 
In such cases it is natural to ask whether gravity may have some influence in the dynamics of phase transitions and, for this reason, the problem of generalizing strongly interacting fermion effective field theories to curved spacetimes triggered a large body of work (See Refs.~\cite{c1,c2,c3,c4,c5,c6,c7,c8,c9,c10,c11,c12,c13,c14,c15,c16} for a partial list of references). 

In curved space, mostly for reasons of technical nature, attention was limited to consider homogeneous spacetimes and condensates, {\it i.e.} spacetime curvature and condensate were assumed not to vary in space. Even in this relatively simple situation, the interesting conclusion was that gravity may, in fact, affect the critical temperature $T_c$ of the theory. Once the theory is immersed in a spacetime of positive constant curvature, in the mean field and large-$N$ approximations, it was shown that, if temperature is fixed to some value $T_<$ smaller than the critical temperature $T_c$, chiral symmetry is broken. Keeping the temperature fixed to $T_<$ and increasing the value of the curvature, produces a phase transition into a chirally restored symmetry phase, indicating a clear analogy between thermodynamical and geometrical effects.

Although this analogy is formally interesting, in concrete physical situations (for example in the case of a neutron star) the value of the curvature is small and geometrical effects only play a marginal role. In such a case, the system can be well described as if the spacetime were effectively flat, limiting the interest of the study of effective strongly interacting theories in curved space. However, this is not a universal situation.

Clearly, geometrical effects become significant in the presence of strong gravitational sources. One natural example that comes to mind is that of black holes. This case, physically very interesting, presents several complications absent in the cases of homogeneous backgrounds. The most important one is related to the fact that approximating the condensate as a spatially constant function is simply not admissible. 

Targeting a solution to this problem, we have recently initiated the analysis of inhomogeneous phases in curved spacetimes. In Ref.~\cite{flachi1}, we developed a formalism that quite naturally allows to deal with this problem in regular ultrastatic spacetimes (the assumption of ultrastaticity was simply motivated by the fact that the special geometric structure allows a direct use of the imaginary time formalism necessary to include finite temperature effects). With some additional efforts, in Ref.~\cite{flachi2}, we have analysed the case of black holes exteriors and shown that chiral phase transitions occur outside the event horizon. More precisely, if the black hole is surrounded by a gas of strongly interacting particles, an inhomogeneous condensate with a kink profile will form. Although the implementation was not straightforward, the idea is, in fact, simple. In a black hole geometry that is asymptotically flat, local quantities increase as the horizon is approached. If the asymptotic temperature, inversely proportional to the black hole mass, is lower than the critical temperature, then chiral symmetry is broken at infinity. According to the same logic, since the local temperature diverges near the horizon, we may expect that, at some finite distance from the horizon, a transition to a chirally symmetric phase may occur. This suggests that for black holes of certain mass in equilibrium with a gas of strongly interacting particles, a bubble of high temperature restored chiral symmetry phase should surround the hole. For an evaporating black hole the same holds, since asymptotically the temperature vanishes as $1/r^2$. This has been quantitatively demonstrated in \cite{flachi2} where self-consistent solutions for condensate have been constructed.

In all cases analysed so far (for both homogeneous and inhomogeneous situations), the spacetime geometry was taken to be regular (in Ref.~\cite{flachi1} we studied the case of regular ultrastatic manifolds, while in Ref.~\cite{flachi2} we focused on the region outside the event horizon that is also singularity free). The goal of this paper is to analyse more deeply the connection between the appearance of inhomogeneous phases and curvature effects, in particular when the geometry presents singularities. 
Examples of this sort describe, for example, global monopoles spaces, geometries with wedges, string theory conifolds. Higher co-dimension brane models also present a similar structure \cite{kaloper}.

The first step of the present work is to compute the effective action for a strongly coupled fermion effective field theory on a conical spacetime. Since the geometry presents singularities, a suitable regularization procedure has to be used to deal with this situation. One possibility is to use the approach developed by Cheeger in Ref.~\cite{cheeger}, where the spectral geometry of differential operators on singular spaces has been discussed. Although the problem can be discussed quite generally, here, for simplicity, let us specify the set-up and consider a quantum field $\chi$ propagating on a $D$-dimensional spacetime $\mathscr{M}$ that presents a conical-type singularity. The effective action can be written in terms of the heat-kernel, $\mathscr{K}$, of some differential operator defined on the spacetime manifold in question. Schematically, we may write
\bea
\mathscr{K}(t)\sim \sum_i \hat{\mathscr{C}}_{i/2} t^{i/2-D/2} + \mathscr{B} \ln t~,
\label{hka}
\eea
where $\hat{\mathscr{C}}_{i/2}$ are the heat-kernel coefficients and the quantity $\mathscr{B}$ is related to the value of the zeta function on the base manifold. In this case, the heat-kernel coefficients are understood as the principal part of integrals over the space manifold of geometric invariants and this is indicated by the hat. Using the above expansion, one may obtain the values $\zeta(0)$ and $\zeta'(0)$ related to the functional determinant of the operator in question. In general the zeta function may present poles and, as shown by Cheeger, by taking the principal part provides the appropriate treatment necessary in obtaining the asymptotic expansion of the heat-trace. In a sense, Cheeger's approach allows to deal directly with the singularity, however, in any practical implementation, truncating in the above expansion becomes necessary, and this affects the validity of the truncated expansion near the singularity.

Another possibility, easier to implement and that we will use here, is to resolve the singularity by excising a small region around it and substituting the excised region with a regular cap (See Fig.~\ref{figg1}) \footnote{A similar type of regularization has also been used in the context of higher co-dimension brane models \cite{kaloper,kaloper2} where the singularity was resolved by placing a brane at small distance from the singularity and by filling the interior by a regular cap.}. Adopting this way of regularization, allows to have a standard expansion for $\mathscr{K}(t)$, {\it i.e.} the coefficient $\mathscr{B}$ vanishes and the heat-kernel coefficients are the ordinary ones. 
\FIGURE[h]{
	\centering
%\put(-7,60){\rotatebox{90}{$\sigma(r)$}}
%\put(105,-5.5){$r$}
	\includegraphics[width=0.5\columnwidth]{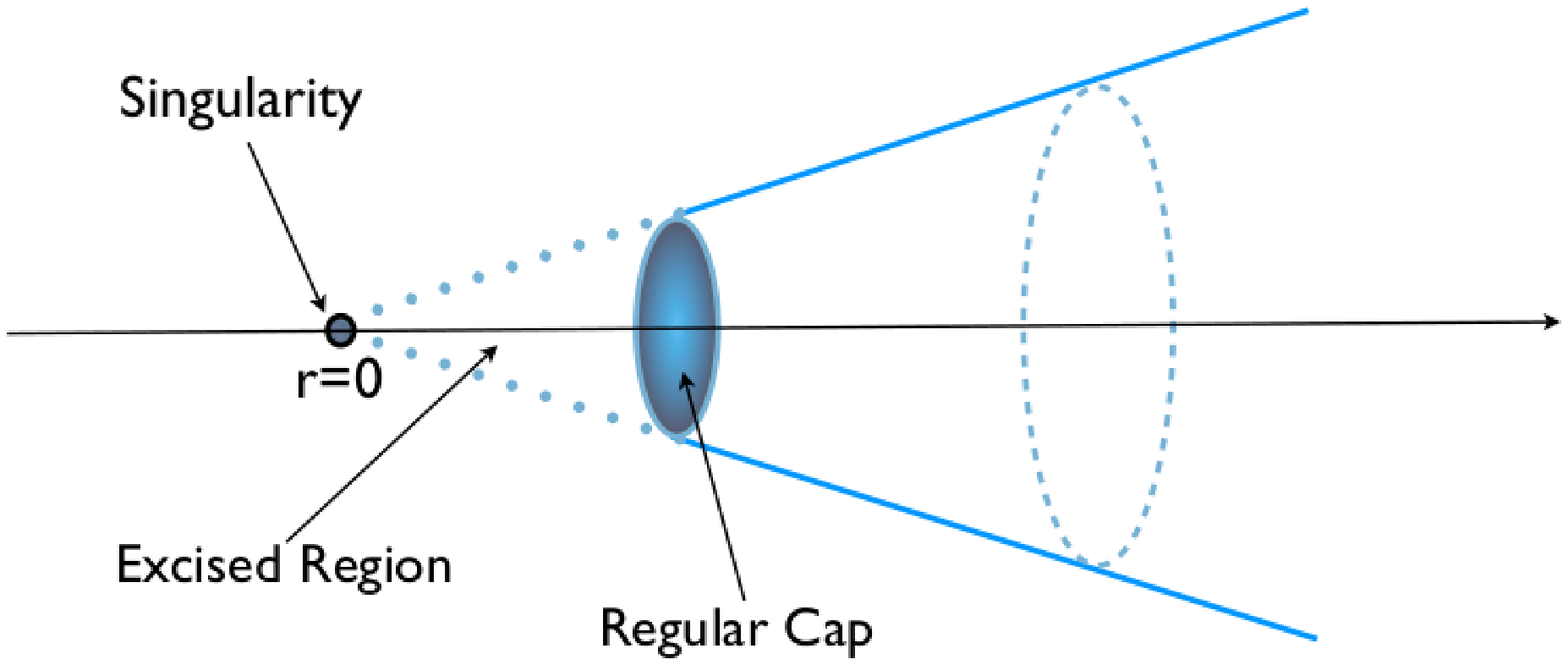}
	\caption{Regularization by excision of the singularity. \vspace{.2cm}}
\label{figg1}
}
Adopting this regularization sche\-me, we will present the computation of the effective action for a strongly interacting fermion effective field theory on manifolds with conifold singularities in the next section.
The method is analogous to that presented in Ref.~\cite{flachi1}.
Once the explicit form of the effective action for the fermion condensate is obtained, we will proceed, in section \ref{sec3} with the numerical implementation and construct self-consistent and spatially non-trivial solutions for the condensate. In particular, we will see that even when the thermodynamical temperature is kept constant and below the critical value for the effective theory in flat space, chiral symmetry is always restored near the singularity. The condensate assumes a kink-like profile that vanishes near the singularity, thus separating the region of restored symmetry around the singularity from a region of broken symmetry. 
Our conclusions will be presented in the last section.

\section{Condensate Effective Action in Conifold-type Geometries}
In this section, we will consider the simplest tractable type of geometric singularity: that of a {\it metric cone}. Given a $(d-1)$-dimensional Riemannian manifold $\mathscr{N}$, a (metric) cone, $\mathscr{C}_{\mathscr{N}}$, is defined as the space $\mathbb{R}^+ \times \mathscr{N}$. In the following we will assume that the base of the cone, $\mathscr{N}$, is compact and smooth. The formal part of the analysis will deal with this general case, while in the subsequent numerical implementation we will specify the form of the base. In local hyperspherical coordinates the line element is
\bea
ds^2 = g_{ij}dx^idx^j=dr^2 +r^2 d\mathscr{N}^2~,
\label{conemetric}
\eea
with $r$ being a radial coordinate, $r \in \mathbb{R}^+$. This is the geometry of a infinitely long cone and it is singular at the apex, $r=0$, where a curvature singularity occurs. It is possible to cut the cone at some distance L, in which case the radial coordinate varies within the interval $I=\left[0,\mbox{L}\right]$. 
%In the following we will keep $L$ large but finite.

Finite temperature effects may be introduced using the standard imaginary time formalism, thus considering the product of $S^1\times \mathscr{C}_{\mathscr{N}}$ where $S^1$ is a circle of radius $2\pi T$, and $T=1/\beta$ being the temperature in natural units. Anti-periodic boundary conditions will be imposed on the fermion fields on $S^1$. Generalization to scalars are almost trivial. 

We will model the strongly interacting fermion effective field theory with an action of the form
\bea
S= \int d^{d+1}x \sqrt{g} \left\{ 
\bar \psi i \gamma^\mu \nabla_\mu \psi 
+ 
{\lambda\over 2N} \left(\bar \psi \psi\right)^2 + \cdots
\right\}~,
\label{action}
\eea
where the spinor $\psi$ has $(4 \times N_f\times N_c)$ components,
$N_f$ and $N_c$ are the number of flavors and colors respectively ($N\equiv N_f \times N_c$), and $\lambda$ is the coupling constant. The dots represent terms with higher mass dimension. The matrices $\gamma_\mu$ are the gamma matrices in curved space, and $g=|\mbox{Det} g_{ij}|$ with $g_{ij}$ given by (\ref{conemetric}). The theory is invariant under discrete chiral transformations, and possible phase transitions in relation to the breaking of the chiral symmetry can be discussed in terms of the appearance of a non vanishing condensate $\sigma = -{\lambda\over N}\langle \bar \psi \psi \rangle$: if the chiral symmetry is broken dynamically, $\sigma$ acquires a non-zero vacuum expectation value and a fermion mass term appears. The basics of the computation are the same as in our previous work Ref.~\cite{flachi1}, so, in recapitulating them, we will be brief.

After bosonization, the effective action at finite temperature (per fermion degree of freedom), $\Gamma$, can be expressed in the large-$N$ approximation as 
\beq
\Gamma= - \int d^dx \sqrt{g} \left({\sigma^2\over 2\lambda}\right) + 
\mbox{Tr} \ln \left( i \gamma^\mu \nabla_\mu - \sigma \right)~,
\label{effect}
\eeq
where the determinant acts on spinor and coordinate space. Squaring the Dirac operator gives
%\begin{widetext}
\bea
\Gamma &=&-\int d^{d}x \sqrt{g} \left({\sigma^2\over 2\lambda}\right) + \delta \Gamma~,
\label{eff}
\eea
with
\bea
\delta \Gamma = {1\over 2}\sum_{\epsilon =\pm}\mbox{Tr} \ln \left[
\square + {R\over 4} +\sigma^2 +\epsilon \left|\partial \sigma\right| 
%+ i \gamma^\mu \sigma_{;\mu}
\right]~,
\label{diff}
\eea
where we have assumed the condensate to be symmetric around the axis of the cone, {\it i.e.} $\sigma=\sigma(r)$. 
Zeta function regularization allows us to express $\delta\Gamma$ as
\bea
\delta \Gamma ={1\over 2}\int d^dx\sqrt{g} \Big(\zeta(0) \ln \ell^2 +\zeta'(0)\Big)~,
\label{dg}
\eea
where $\ell$ is a renormalization scale. The quantities $\zeta(0)$ and $\zeta'(0)$ are the analytically continued values to $s=0$ of the following function
\bea
\zeta(s) = {1\over \Gamma(s)} \sum_n \sum_\epsilon \int dt t^{s-1} e^{-t\omega_n^2} \mbox{Tr} e^{-t\left(-\Delta + {R\over 4} +\sigma^2 +\epsilon \left|\partial \sigma\right| \right)}~.
\eea
In the above expression we have used the imaginary time formalism to introduce finite temperature effects ($\omega_n= 2\pi(n+1/2)/\beta$ are the Matsubara frequencies for the fermion fields), and we expressed the zeta function in terms of the Mellin transform of the heat-trace. 

In the following we will regularize the effective action by excising the singularity as this seems the most convenient way to deal with the subsequent numerical analysis. In this case the heat-kernel expansion will take the standard form and the radial coordinate is assumed to start from $r=\epsilon$, where boundary conditions will be imposed. Some algebra allows us to recast the above expression as
\bea
\zeta(s)= \zeta_+(s) + \zeta_-(s)~,
%+\zeta^{sing}_+(s)+\zeta^{sing}_-(s)~,
\label{zpm}
\eea
where  
\bea
\zeta_\pm(s) = \int dt {t^{s-1} \over \Gamma(s)} {e^{-t \tilde\sigma^2_\pm} 
\mathscr{K}_{reg}(t)\over 2\sqrt{\pi t}} g(t)~,
\nonumber
\label{zeta}
\eea
with
\bea
g(t)&=&\left(1+2\sum_{n=1}^\infty (-1)^n e^{-{\beta^2n^2\over 4t}}\right)~,
\\
\tilde \sigma^2_\pm &=& R/12 +\sigma^2 \pm \left|\sigma'\right|~,
\eea
Notice that in the regular part of the heat-trace we have factorized out the exponential according to the method described in Refs.~\cite{toms,jack}. 
%This allows to perform a partial resummation of the heat-kernel and to slightly simplify the expression of the coefficients. Details of this procedure relevant to our case have discussed in Ref.~\cite{flachi1} and we defer the reader to that reference for details. 
In the above expression $s$ is regulating parameter assumed to lie in a region of the complex plane where all the above integrals are well defined. The standard expression for $\mathscr{K}_{reg}(t)$ is given by
\bea
\mathscr{K}_{reg}(t) &=& {1\over (4\pi)^{d/2}}\sum_{k=0}^\infty \mathscr{C}_\pm^{(k/2)} t^{k/2-d/2}~,
\label{kreg}
\eea
where the coefficients ${\mathscr{C}}^{(k/2)}_\pm$ are the heat-kernel coefficients associated with the differential operator in (\ref{effect}) and are integrals of polynomials in the curvatures on the regularized spacetime manifold and therefore are all regular. Coefficient with integer index refer to volume contributions while coefficients with semi-integer indices to boundary contributions. The first coefficient is $\mathscr{C}_\pm^{(0)}=1$, while $\mathscr{C}_\pm^{(1)}=0$. Higher order coefficients will be given later as needed.

Using the above expression for $\zeta_\pm(s)$, we can express the effective action as
\bea
\Gamma &=&-\int d^{3}x \sqrt{g} \left({\sigma^2\over 2\tilde g}\right) + \Gamma_+ +\Gamma_- 
%+ \Gamma_{+}^{sing}+ \Gamma_{-}^{sing}
~,
\eea
where $\Gamma_\pm$ are the contributions to the effective action coming from $\zeta_\pm$. 
%and $\Gamma_{\pm}^{sing}$ the one coming from $\zeta_{sing}$ composed according to formulas (\ref{dg}) and (\ref{zpm}).
For completeness, at the end of this section, we will illustrate how the computation can be performed also by using the full heat kernel expansion including the logarithmically singular term in (\ref{hka}).

We will start assuming $s$ to be in the region where the integrals are well defined, then analytically continue to $s=0$ and take the limit $d=3$.
The zeta function can be expressed in terms of the following integral and its derivative with respect to the regularizing parameter $s$,
\bea
\mathscr{I}(a,b,c) &=& {1\over (4\pi)^{d+1\over 2}} \int {dt\over \Gamma(s)}\, t^{s-a}\,e^{{-c t }-b/t}~,\label{I}~.
\eea
These can be computed explicitly,
%The interested reader may consult chapter 20 of the book \cite{lang} for a more rigorous treatment of these functions.
using the following expressions (see Ref.~\cite{lang}):
\bea
\mathscr{I}(a,b,c) &=&
{2^{a-s}\over (4\pi)^{d+1\over 2}} 
{1\over \Gamma(s)}
\left({c\over b}\right)^{(-1+a-s)/2}K_{a-s-1} \left(2\sqrt{b c}\right)~,\nonumber
\\
{d \mathscr{I}(a,b,c) \over ds} &=&
{1\over \Gamma(s)} {1\over (4\pi)^{d+1\over 2}}
\left({c\over b}\right)^{(-1+a-s)/2}
\left[
-K_{a-s-1}\left(2\sqrt{b c}\right) \left(
\ln\left({c\over b}\right)
+2 \psi^{(0)}(s) \right)\right.\nonumber\\
&&\left.-2 \left.{d\over d\nu}K_{\nu}\left(2\sqrt{b c}\right)\right|_{\nu=a-s-1}
\right]~,\nonumber
\eea
The above expression are valid under the following conditions:
\bea
&&\Re b > 0~,\nonumber\\
&&\Re c > 0~,\nonumber\\
&&\Re \left( a -s \right) >1~.
\eea
The first two requirements are trivially satisfied in our case. The third is instead dealt with in the usual way of zeta function regularization, {\it i.e.} by assuming that $s$ lies in a region of the complex plane where the third inequality is satisfied and then proceed by analytical continuation.  One may easily re-write the zeta function above as
\bea
\zeta(s) = \sum_{\epsilon=\pm} \sum_{i=1}^2 \mathscr{Z}_\epsilon^{(i)}(s)~, 
\label{zf}
\eea
where we have defined for notational convenience
\bea
\mathscr{Z}^{(1)}_\pm(s) &=&  \sum_{k=0}^\infty \mathscr{C}_\pm^{(k/2)} 
\mathscr{I}({3+d-k\over 2}, 0,\tilde\sigma^2_\pm)~,\label{z1}\\
%\int dt\, t^{s-3/2+k/2-d/2} e^{-t \tilde\sigma^2_\pm} 
\mathscr{Z}^{(2)}_\pm(s) &=&  2 \sum_{k=0}^\infty \sum_{n=1}^\infty (-1)^n \mathscr{C}_\pm^{(k/2)} 
\mathscr{I}({3+d-k\over 2},{\beta^2n^2\over 4},\tilde\sigma^2_\pm)~,\label{z2}
\eea

Dividing now $\Gamma_\pm$ as the sum of the contribution coming from $\zeta(0)$, $\Gamma_\pm^{(0)}$, and that coming from $\zeta'(0)$, $\Gamma_\pm^{(1)}$, allows us to write
\bea
\Gamma_\pm =  \Gamma_\pm^{(0)} + \Gamma_\pm^{(1)}~. 
\eea
A direct computation shows that only $\mathscr{Z}_\pm^{(0)}$(s) gives a non vanishing contribution to $\zeta(0)$. We find
\bea
\Gamma_\pm^{(0)} = {1\over 32\pi^2} \int dr\,r^2\,\sqrt{\mbox{det}_{\mathscr{N}}}\,
\left[
{1\over 2}\tilde\sigma_\pm^4 -\tilde\sigma_\pm^2\mathscr{C}^{(1)}_\pm
+\mathscr{C}^{(2)}_\pm
\right]\ln \ell^2,~
\label{g0}
\eea
where $\det_{\mathscr{N}}$ is the determinant of the metric on the base manifold $\mathscr{N}$. The above result is exact (no truncation in the heat-kernel expansion has been done and boundary contributions also vanish). %The reader may easily check with no effort that the integrals in the above expression are regular. 
Also, notice that the ansatz used in the regular part of the heat-trace gives $\mathscr{C}^{(1)}_\pm =0$ (See Refs.~\cite{toms,jack,flachi1}). Therefore the second term in the above expression vanishes. The coefficient $\mathscr{C}^{(2)}_\pm$ is given by
\bea
\mathscr{C}^{(2)}_{\lambda} &=& {1\over 180}R_{\mu\nu\rho\sigma} R^{\mu\nu\rho\sigma} -{1\over 180}R_{\mu\nu} R^{\mu\nu}- {1\over 120} \Delta R + {1\over 3} \left(\left(\sigma\pm {1\over r}\right)\sigma''+\sigma^{'2}+{2\over r}\sigma \sigma'\right)~,\nonumber
\eea
where
the quantities $R$, $R_{\mu\nu}$ and $R_{\mu\nu\lambda\rho}$ are the Ricci scalar and Ricci and Riemann tensors, respectively, for the geometry (\ref{conemetric}). All terms proportional to curvature invariants and not to the condensate will not, in fact, contribute to effective action for the condensate, in the sense that they will disappear in the equation of motion for $\sigma$. The integrand in (\ref{g0}), {\it i.e.} the contribution to the Lagrangian density for the condensate, is regular in the limit $r\rightarrow 0$.

The term that requires more work is the contribution to the effective action that comes from the derivative of the zeta function, $\Gamma_\pm^{(1)}$. Proceeding as described above, we find for the regularized effective action the following expression
\bea
\Gamma^{(1)}_\pm &=& 
{1\over 32\pi^2} \int dr\,r^2\,\sqrt{\mbox{det}_{\mathscr{N}}}\,
\Bigg\{
{3\over 4} \tilde\sigma^4_\pm
-\left({1\over 2}\tilde\sigma^4_\pm +\mathscr{C}^{(2)}_\pm\right)\ln \tilde \sigma^2_\pm
\nonumber\\
&&
+4\sum_{n=1}^{\infty}(-1)^{n}
\Bigg[
{4\tilde \sigma^2_\pm \over n^2 \beta^2} K_2\left(n\beta\tilde \sigma_\pm\right)
+\mathscr{C}^{(2)}_\pm
K_0\left(n\beta\tilde \sigma_\pm\right)
\Bigg] +\cdots \Bigg\}~.
\label{eapm}
%\\
\eea
The dots represent higher order terms in the heat kernel expansions. They can be computed explicitly, but we won't report their form here. All higher order terms give contributions in the derivatives of the condensate of order equal or greater than four, so the expression above is sufficient when keeping the analysis to second order in the derivative of the condensate. Boundary terms can also be calculated easily and the first few terms are
\bea
%&&
\Gamma_{boundary}&=&{1\over 32\pi^2} \int d^3x \sqrt{g}\,\delta(r-L)\,
\Bigg\{{4\over 3}\sqrt{\pi}\tilde\sigma^3_\pm
\mathscr{C}^{(1/2)}_\pm
-2\sqrt{\pi} \tilde \sigma_\pm \mathscr{C}^{(3/2)}_\pm
\nonumber\\
&&
+4\sum_{n=1}^{\infty}(-1)^{n}
\Bigg[
+{2\sqrt{\pi}\over n^2\beta^2} \mathscr{C}^{(1/2)}_\pm
\left(\tilde \sigma_\pm+{1\over n\beta}\right)
e^{-n\beta\tilde \sigma_\pm}
+{\sqrt{\pi}\over n\beta} \mathscr{C}^{(3/2)}_\pm e^{-n\beta\tilde \sigma_\pm}
\Bigg]+\cdots \Bigg\},~
\nonumber
\eea
where the dots represents, as before, higher order terms in the heat-kernel expansion. 
As before, one can easily verify the regularity of the expressions above. 

To conclude this section, we will show how the full heat-kernel expansion (\ref{hka}), including the logarithmic terms, can be used instead of the scheme adopted above. In this case, according to (\ref{hka}) the heat-kernel will consist of two terms. The polynomial part of the heat-kernel can be treated as above with the only difference being that the heat-kernel coefficients are understood as the principal part of integrals over the space manifold. Since the manifold has singularities, the integration in the heat-kernel coefficients, naturally, leads to divergences. As shown in Ref.~\cite{cheeger}, the relevant quantity for the heat-kernel asymptotics is the principal part of these diverging integrals that can be extracted in several ways. One possibility is to cut the range of integration at finite distance from the singularity and this is equivalent to the regularization by excision used above. The additional (logarithmic) term in (\ref{hka}) can also be included in the analysis, but some attention is necessary. The quantity $\mathscr{B}$ is proportional to minus the residue of the zeta function at $s=0$, $-\mbox{Res}\,\zeta_{\mathscr{C}_{\mathscr{N}}}(0)$, that can, in turn, be related to the value of the zeta function evaluated on the base manifold analytically continued to $s=-1/2$.
%\bea
%\mathscr{B} = - = -{1\over 2} \mbox{Res}\,\zeta_{\mathscr{N}}(-1/2)~.
%\eea
The logarithmic part of the heat-kernel expansion will lead to an additional term in zeta function that can be written as 
\bea
\zeta(s) = \sum_{\epsilon=\pm} \sum_{i=3}^4 \mathscr{Z}_\epsilon^{(i)}(s)~, 
\eea
with
\bea
\mathscr{Z}^{(3)}_\pm(s) &=&  {\mathscr{B}_\pm\over \sqrt{4\pi}} 
\mathscr{H}_\pm(3/2,0,0)~,\label{z3}\\
%\int dt t^{s-3/2}\, \ln t\,e^{-t \tilde\sigma^2_\pm} 
\mathscr{Z}^{(4)}_\pm(s) &=&  2 {\mathscr{B}_\pm\over \sqrt{4\pi}} 
\sum_{n=1}^\infty (-1)^n 
\mathscr{H}(3/2,{\beta^2n^2\over 4},0)~,\label{z4}
%\int dt t^{s-3/2}\,\ln t\,e^{-t \tilde\sigma^2_\pm-{\beta^2n^2\over 4t}} 
\eea
where
\bea
\mathscr{H}(a,b,c) &=& {1\over (4\pi)^{d+1\over 2}} \int {dt\over \Gamma(s)}\,t^{s-a}\,\ln t\, e^{-c t-
b/t}\label{H}~.
\eea
The analytical continuation of the logarithmic term can be done straightforwardly by using the following expressions
\bea
\mathscr{H}(a,b) &=& 
{b^{(1-a+s)/2}\over (4\pi)^{d+1\over 2}} {\Gamma(-1+a-s)\over \Gamma(s)} 
\left( \psi^{(0)}\left(-1+a-s\right) -\ln b \right) ~,\nonumber
\\
{d \mathscr{H}(a,b)\over ds} 
&=& 
{b^{(1-a+s)/2}\over (4\pi)^{d+1\over 2}}
{\Gamma(-1+a-s)\over \Gamma(s)} 
\left[
\left( \psi^{(0)}\left(-1+a-s\right) -\ln b \right)\times \right.\nonumber\\
&&
\left.\left( \psi^{(0)}\left(-1+a-s\right) +\psi^{(0)}(s) -\ln b \right)
+ \psi^{(1)}\left(-1+a-s\right) \right]
~,\nonumber
\eea
whose analytical continuation can be performed as before. 
A straightforward computation shows that the contribution of the logarithmic term to $\zeta(0)$ vanishes, while the contribution to the effective action from $\zeta'(0)$ gives
\bea
\Gamma^{log}_\pm &=& =
{\ln 2 \over (4\pi)^{d\over 2}\beta}\left(\gamma_e -\ln\left(2\beta^2\right)\right)
\int d^dx \sqrt{g}
{\mathscr{B}_\pm} ~.
\label{zsing}
\eea
%When the base manifold is a positively curved space with constant curvature (for example a sphere) the value of the above zeta function is a constant when the associated differential operator is the Laplacian. This is not the case here, since the operator depends on the condensate. Assuming that the condensate is a regular constant function near the singularity implies that $\mathscr{B}_\pm$ approaches a constant value there. Then the integration over $r$ becomes trivally convergent for $d=3$. The assumption is necessary here only to show that the above contribution to the effective action is regular and it can be verified {\it a posteriori} after solving self-consistently for the condensate. 

%For the convenience of the reader we will report the computation of the contribution from the singular term in a slightly different way. First of all, one may notice by direct computation that the only term that contribute to the effective action is 
%\bea
%&&\sum_n (-1)^n {d\over ds} \int {dt\over \Gamma(s)} t^{s-3/2 \ln t\, e^{-\beta^2 n^2/(4t)}} \nonumber\\
%&=& \sum_n (-1)^n {2\sqrt{\pi} \over n\beta}\left( \gamma_e +2 \ln(n \beta)\right)\nonumber \\
%&=& {2\sqrt{\pi} \over \beta}\left( \gamma_e +2 \ln(2\beta^2)\right)~. 
%\eea

\section{Inhomogeneous Condensates in Monopole Geometries: Numerical Construction}
\label{sec3}
This section will be devoted to implement numerically the above formalism and to construct explicit solutions for the condensate. To have a concrete working model, we need to specify the geometry of the base. In order to make contact with a case of physical interest, we will choose the base manifold to be a sphere of non-unitary radius $\rho$. The curvature is $R\propto(1-\rho^2)\rho^{-2}r^{-2}$, the space is non flat when $\rho\neq 1$. In the following we will consider $\rho<1$, therefore $R>0$. In this case, the metric (\ref{conemetric}) describes the spatial section of global monopole \cite{vilen}. For this geometry, the heat-kernel analysis of the conformal Laplacian has been done, for instance, in Ref.~\cite{bordag}.

The equation of motion for the condensate can be found by minimizing the effective action, that is by varying $\Gamma$ with respect to the condensate $\sigma$. In the previous section we have obtained the effective action to fourth order in the heat-kernel expansion ({\it i.e.} $k=4$ in (\ref{kreg})), leading to a second order non-linear equation of motion for the condensate. This can be written, with some work, in the form of non-linear Schr\"odinger-like equation, whose explicit form is very lengthly and will be omitted here.

The structure of the solution may be anticipated by looking at the form of the thermodynamic potential. At large distance from the singularity, where curvature effects are negligible, the thermodynamic potential ${U_{as}}(\sigma)$ is expected to have a behaviour similar to flat space: below (above) some critical value of the temperature, chiral symmetry is broken (restored). This can be verified by letting $\sigma'=0$, taking the limit $r\rightarrow \infty$ of the effective potential, and by computing the thermodynamic potential by numerical integration of $\partial_\sigma {U}$ with respect to $\sigma$. The profile of the asymptotic potential is shown in Fig.~\ref{fig1} (left panel). 
The thermodynamic potential can also be computed locally and the result is illustrated in Fig.~\ref{fig1} (right panel). The figure shows that even if chiral symmetry is broken asymptotically (right-most (purple) curve), chiral symmetry will gradually be restored as we move towards the singularity, {\it i.e.} the minima of the potential will gradually shift towards a configuration with vanishing $\sigma$. 
\FIGURE[h]{
	\centering
\put(-7.,60){\rotatebox{90}{$U_{as}(\sigma)$}}
\put(100,-5.5){$\sigma$}
\put(310,-4){$\sigma$}
\put(410,65){\rotatebox{90}{$U(\sigma)$}}
	\includegraphics[width=0.47\columnwidth]{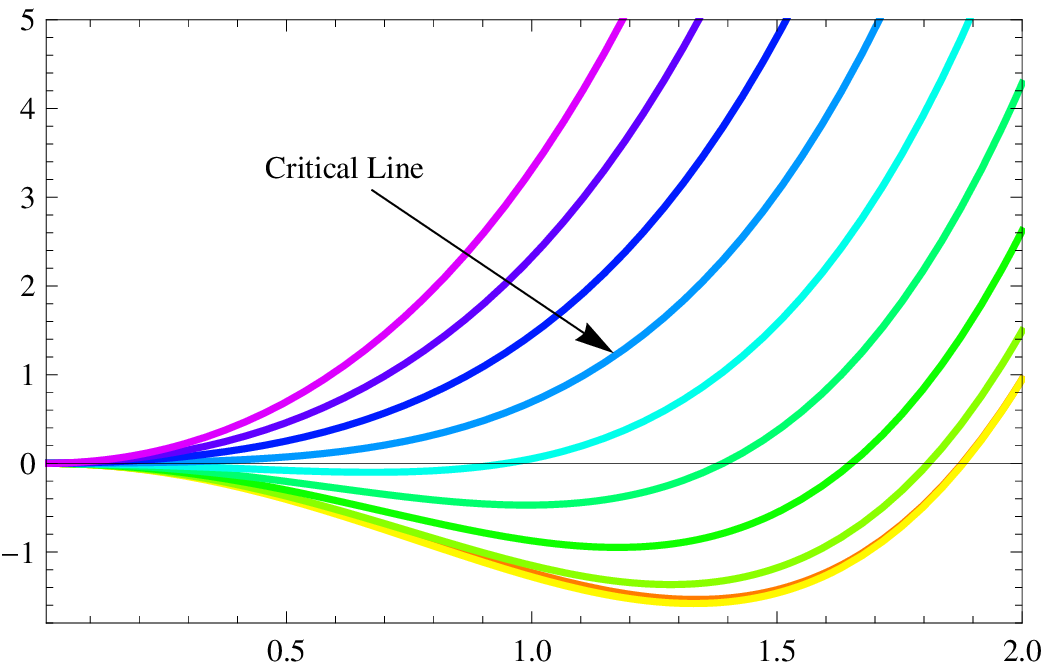}
	\includegraphics[width=0.47\columnwidth]{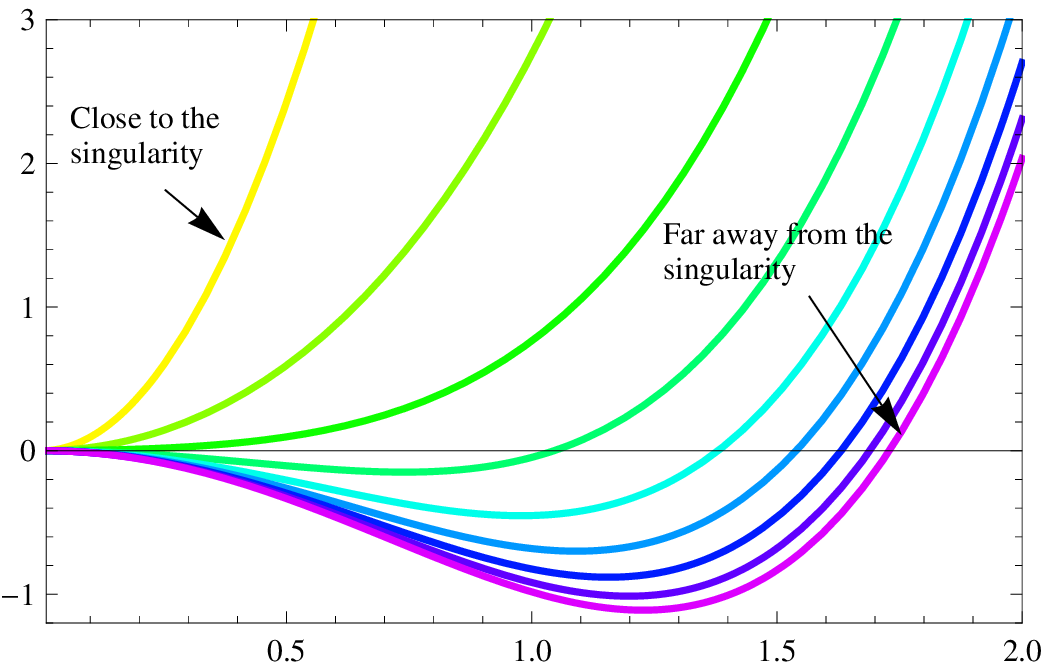}
	\caption{The figure on the left illustrates how the local thermodynamic potential changes as the temperature increases. The right-most (left-most) curve shows the potential for temperature smaller (higher) than the critical one. The figure on the right illustrates how the local thermodynamic potential changes as the singularity is approached. The right-most curve shows the potential at large distance from the singularity for a set of parameters that correspond to a chirally broken symmetry phase (The minima of the potential is non vanishing). The left-most curve shows the potential close to the singularity ($r=0.1$) for which the potential has a vanishing minima at $\sigma=0$, leading to a chirally restored symmetry phase. For both figures we set $\ell=10^{6}$ and $\lambda=10^{-2}$.}
\label{fig1}
}
The solution for the condensate can be found using standard numerical methods, and here we have used a fourth order Runge-Kutta one. In solving the equation for the condensate numerically, there are several things that require some care. One has to deal with the infinite summations over Bessel functions with argument proportional to the condensate. The present situation differs from the black hole case studied earlier \cite{flachi2} as described in the following. At large distance curvature effects are negligble and the argument of the Bessel function is not small. In this case, one may conveniently truncate the summation due to the fact that the Bessel function decay exponentially for large arguments. Near the singularity the contribution from the curvature is, instead, large, and the argument of the Bessel functions is again not small. In the intermediate region for $r$ sufficiently large, we require that, when the argument of the Bessel functions becomes small, the summands are expanded for small arguments and full resummation over $n$ is performed. We then proceed by matching the solutions in the truncated and resummed domains. With the equations of motion for the condensate explicitly written, we solve them by requiring regularity for the solution. The boundary conditions on the solution are set numerically by requiring, near the singularity, that the condensate is at a minima of the potential and by fine tuning the value of the derivative to match the asymptotic value of the solution with the minima of the potential at infinity. 
\FIGURE[h]{
	\centering
\put(-7,60){\rotatebox{90}{$\sigma(r)$}}
\put(105,-5.5){$r$}
	\includegraphics[width=0.5\columnwidth]{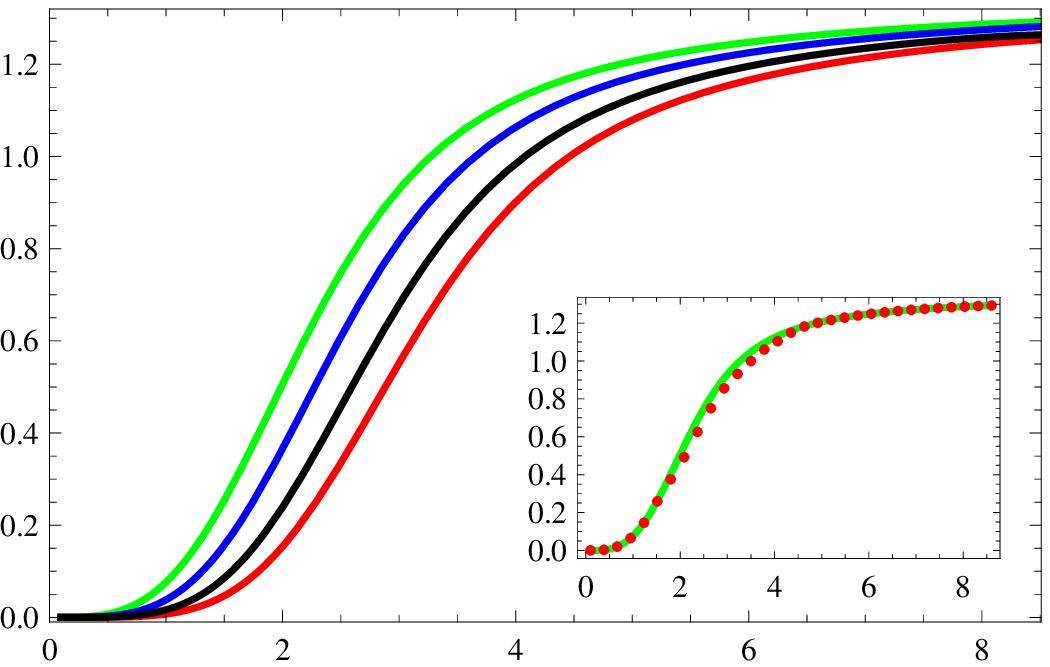}
	\caption{Condensate profile found by numerically minimizing the effective action (\ref{effect}), for four indicative values of the base manifold radius (Left to right: $\rho=0.65~\mbox{(Green)},0.57~\mbox{(Blue)},0.53~\mbox{(Black)},0.48~\mbox{(Red)}$). We set to $\ell=10^6$, $\lambda=10^{-2}$. 
The superposed panel shows the solution for the condensate when higher order terms included superposed over the background solution. The continuous line refers to the background solution corresponding to $\rho=0.65\,\mbox{(Green)}$ and the dotted (red) solution corresponds to the higher order one. \vspace{.2cm}}
\label{fig2}
}
The results for the condensate profile are illustrated in Fig.~\ref{fig2} for sample values of the parameters showing that near the singularity the condensate vanishes and chiral symmetry is restored, while, as we gradually move away from the singularity, the condensate assumes a non vanishing expectation value and chiral symmetry breaks. Notice that contrary to the black hole case we have studied earlier \cite{flachi2}, in the present case, the thermodynamical temperature is constant uniformly along the spatial section of the space and the effect of breaking/restoration of the chiral symmetry is solely due to genuine curvature effects.
As in the black hole case \cite{flachi2}, this indicates that a bubble of chirally restored symetry phase surrounds the singularity.\\
The analysis can be extended to higher orders by including higher order terms in the heat-kernel expansion. The algebra becomes rather cumbersome, but it is possible to handle, with some work, all the computations automatically by using any computer algebra program. After including terms up to fourth order derivative of the condensate, we implement our analysis by perturbing the background solution, $\sigma = \sigma_{bg}+\delta \sigma$, and expanding the higher order equation for $\sigma$ keeping terms up to second derivative of the perturbation. Proceeding in this way, we are assuming that the corrections due to higher order terms only produce small changes in the background solution. This assumption will be verified {\it a posteriori} after the solution is obtained. Once higher order derivatives of the perturbation are dropped out, the equation for the $\delta \sigma$ becomes again a second order one with source term and can be solved by standard methods. This approach is similar to the case of black holes we have addressed earlier \cite{flachi2} and, as in that case, the corrrection to the solution only produces very small distortions that decrease with the thickness of the kink. A sample plot is shown in the small panel in Fig.~\ref{fig2}.

\section{Conclusions}

In this paper we have continued the analysis of the chiral symmetry breaking of strongly coupled fermion effective field theory in curved space, extending our previous work to include conifold-type geometries.

The results, in agreement with our intuition, show that, near the singularity, the condensate vanishes and chiral symmetry is restored. As we gradually move away from the singularity, the condensate assumes a non vanishing expectation value and chiral symmetry breaks. This indicates that the singularity will be surrounded by a bubble of chirally restored symmetry phase, in much the same way as black holes (See Ref.~\cite{flachi2}). We have illustrated this explicitly in the tractable example of a conifold geometry with generic base manifold (the generalized metric cone). More precisely, we have considered a strongly interacting fermion effective field theory propagating on the conifold and constructed an explicit solution for the condensate. The formal analysis has been carried out in general, and we have evaluated the effective action for the condensate introducing finite temperature effects by means of the standard Matsubara formalism. This set-up allowed us to keep the genuine thermodynamical temperature constant and to clearly single out any non-trivial effect of the curvature. 
The evaluation of the effective action was carried out using a sophistication of the method we have described in Ref.~\cite{flachi1} and that makes use of zeta function regularization. Due to the presence of the singularity the regularization requires some care, however zeta function regularization proves to be very adequate in this case.

Several generalizations may be considered. First of all, one expects that the present situation is more general. That is in the vicinity of {\it any} spacetime singularity the same occurs. This can in principle be treated by studying the heat-kernel asymptotics in spacetimes with different singular structure. Although computational complications may arise, the procedure presented in this paper that regularizes the geometry by excising a small region around the singularity should work with little modifications. Generalizations to more sophisticated models may also be possible with some effort, the most interesting ones being the inclusion of gauge degrees of freedom and of a chemical potential.

It was our goal to understand how condensates and chiral symmetry may be affected by curvature singularities and for this reason we considered a purely geometrical background (\ref{conemetric}). However, the results may be easily extended to cases where a non-trivial radial dependence of the lapse function occurs, and this can be handled by using the same conformal techniques that we have adopted to study the case of black holes. In this case, if the local temperature increases as the singularity is approached we expect the same phenomena of chiral restoration described here to occur.

\acknowledgments
The financial support of the Funda\c{c}\~{a}o p\^{a}ra a  Ciencia e a Tecnologia of Portugal (FCT) is gratefully acknowledged. I wish to express my gratitude to Takahiro Tanaka for his crucial help with several technical points, for carefully reading the manuscript and for many useful suggestions that helped to improve the presentation of the material of this paper. I also wish to thank Marco Ruggieri for continuous discussions on strongly interacting fermion effective field theories.

\end{document}